\documentclass[aps,prl,twocolumn,preprintnumbers,nofootinbib]{revtex4-1}
\usepackage{amsmath, mathrsfs, amssymb,amsfonts,amsthm,graphicx, epsf, dcolumn, yfonts}
\usepackage[utf8]{inputenc}
\usepackage[hyperfootnotes=true]{hyperref}
\usepackage{color}
\usepackage{slashed}
\usepackage{setspace}
\usepackage{cancel}
\usepackage{wasysym}
\usepackage[lofdepth,lotdepth]{subfig}
\usepackage{float}
 
\pdfoutput=1
 \newcommand{\be}{\begin{equation}}
 \newcommand{\ee}{\end{equation}}
 \newcommand{\bea}{\begin{eqnarray}}
 \newcommand{\eea}{\end{eqnarray}}

\newcommand{\beq}{\begin{equation}}
\newcommand{\eeq}{\end{equation}}

\begin{document}


\title{Microcanonical action and the entropy of Hawking radiation} 


\author{Juan F. Pedraza,$^{1}$ Andrew Svesko,$^{2}$ Watse Sybesma$^{3}$ and Manus R. Visser$^{4}$}
\affiliation{$^{1}$Departament de Física Quàntica i Astrofísica and
  Institut de Ciències del Cosmos, Universitat de Barcelona, 08028 Barcelona, Spain\\
$^2$Department of Physics and Astronomy, University College London, London WC1E 6BT, United Kingdom\\
$^3$Science Institute, University of Iceland,\\
Dunhaga 3, 107 Reykjav\'ik, Iceland\\
$^4$Department of Theoretical Physics, University of Geneva,\\
24 quai Ernest-Ansermet, 1211 Gen\`{e}ve 4, Switzerland
}

\begin{abstract}\vspace{-2mm} 
\noindent

\noindent The island formula -- an extremization prescription for  generalized entropy -- is known to result in a unitary Page curve for the entropy of Hawking radiation. This semiclassical entropy formula has been    derived for Jackiw-Teitelboim (JT) gravity coupled to conformal matter using the ``replica trick" to evaluate the Euclidean path integral. Alternatively, for eternal Anti-de Sitter black holes, we derive the extremization of generalized entropy from  minimizing the microcanonical action of an entanglement wedge. The   on-shell action is minus the entropy and   arises in the saddle-point approximation of the (nonreplicated) microcanonical path integral. When the black hole is coupled to a bath, islands emerge from maximizing the   entropy   at fixed energy, consistent with the island formula. Our method applies to JT  gravity  as well as other two-dimensional dilaton gravity theories.

\end{abstract}

\renewcommand*{\thefootnote}{\arabic{footnote}}
\setcounter{footnote}{0}

\maketitle

 \setcounter{secnumdepth}{1}

\section{Introduction}

 Hawking's discovery \cite{Hawking:1974rv,Hawking:1974sw} of thermal radiation emitted by black holes has led to a puzzle quantum gravity is expected to resolve: the information paradox~\cite{Hawking:1976ra}. A solution will provide a better understanding of the quantum mechanical evolution of black holes. Either black holes evolve unitarily, such that the von Neumann (vN) entropy of radiation $S_{\text{vN}}^{\text{rad}}$ follows a Page curve \cite{Page:1993wv,Page:2013dx}, or they do not. For  dynamical black holes  formed from gravitational collapse,  the Page curve has the generic feature  that $S_{\text{vN}}^{\text{rad}}$ increases from zero (as the radiation begins in a pure state), until it reaches a maximum at the Page time $t_{\text{P}}$,  and then decreases to zero at late times, thus conserving information. Advances motivated by the Anti-de Sitter/Conformal Field Theory (AdS/CFT) correspondence suggest $S_{\text{vN}}^{\text{rad}}$ indeed evolves unitarily \cite{Penington:2019npb,Almheiri:2019psf,Almheiri:2019yqk,Almheiri:2019hni,Almheiri:2019qdq,Penington:2019kki,Goto:2020wnk,Hollowood:2020kvk,Hollowood:2020cou,Wang:2021mqq}.

 Surprisingly, the unitary Page curve can be derived within semiclassical gravity. Crucial to this derivation are quantum extremal surfaces (QESs) -- codimension-2 surfaces $X$ which extremize the semiclassical generalized entropy $S_{\text{gen}}$ -- and the QES formula 
 \cite{Hubeny:2007xt,Faulkner:2013ana,Engelhardt:2014gca} 
\beq S_{\text{vN}}(\Sigma_X)=\underset{X}{\text{min}}\,\underset{X}{\text{ext}}\left[\frac{\text{Area}(X)}{4G}+S^{\text{sc}}_{\text{vN}}(\Sigma_{X})\right] .\label{eq:QESformula}\eeq
Here, $S_{\text{vN}}(\Sigma_X)$ is the fine-grained vN entropy of $\Sigma_X$ in the full quantum theory, $S^{\text{sc}}_{\text{vN}}$ is the vN entropy of bulk quantum fields including   both  matter fields and the metric in the semiclassical approximation, and $\Sigma_{X}$ is a codimension-1 slice bounded by $X$ and a cutoff surface. The bracketed 
term is the generalized entropy
 $S_{\text{gen}}(\Sigma_X)$, 
 and obeys a generalized second law \cite{Bekenstein:1974ax,Wall:2010cj,Wall:2011hj}.

The QES formula (\ref{eq:QESformula}) also holds for the vN entropy of Hawking radiation $S_{\text{vN}}^{\text{rad}}$, where it is known as the ``island formula" \cite{Almheiri:2019hni}. Indeed, applying  (\ref{eq:QESformula}) to evaporating or eternal black holes reveals a Page curve \cite{Penington:2019npb,Almheiri:2019psf,Almheiri:2019yqk}.  In this case $\Sigma_{X}$ may be disconnected, $\Sigma_{X}=\Sigma_{\text{rad}}\cup I$,
where $\Sigma_{\text{rad}}$ is the region outside the black hole 
and $I$ is an ``island" with $X=\partial I$. For evaporating black holes $I$ lies inside the black hole   \cite{Almheiri:2019hni,Hollowood:2020cou,Wang:2021mqq}, while for eternal black holes islands extend outside the classical horizon \cite{Almheiri:2019yqk,Gautason:2020tmk,Hartman:2020swn,Chen:2020hmv,He:2021mst}. 
Motivated by \cite{Lewkowycz:2013nqa,Faulkner:2013ana,Dong:2016hjy,Dong:2016fnf,Dong:2017xht}, the island formula  has been derived using the ``replica trick" in the context of Jackiw-Teitelboim (JT) gravity \cite{Almheiri:2019qdq,Penington:2019kki,Goto:2020wnk}. 
The Page curve  arises from a   competition between two saddle point geometries dominating the Euclidean gravitational path integral (PI). At early times the (replicated)   black hole solution, or ``Hawking saddle'', dominates the PI and is responsible for the rise in $S_{\text{vN}}^{\text{rad}}$. At late times, ``replica wormholes" overtake the black hole leading to the ``island saddle" and, for evaporating black holes, a decrease in $S_{\text{vN}}^{\text{rad}}$.

Working in the microcanonical ensemble   \cite{Brown:1992bq,Banados:1993qp}, here we derive the extremization condition in (\ref{eq:QESformula})    from a (nonreplicated) $\text{AdS}_{2}$ gravitational PI. Indeed, for  black holes with  $U(1)$ Killing symmetry, the replica trick is not necessary to compute gravitational fine-grained entropies \cite{Casini:2011kv}. 
Rather, one may opt to use the standard thermal gravitational partition function \cite{Gibbons:1976ue}. Further,  in this ensemble we show how   islands arise for eternal $\text{AdS}_{2}$ black holes coupled to a bath, a setting with its own information paradox~\cite{Almheiri:2019yqk}.
 For definiteness, we will focus on semiclassical JT gravity but our methodology applies to more general two-dimensional (2D) dilaton theories of gravity.

\section{Semiclassical JT gravity and the Wald entropy}

 Semiclassical JT gravity is characterized by the action 
\beq I =I_{\text{JT}}+I_{\text{Pol}}\label{eq:totscact}\,,\eeq
where $I_{\text{JT}}=I_{\text{JT}}^{\text{bulk}}+I_{\text{JT}}^{\text{GHY}}+I_{\text{JT}}^{\text{ct}}$ is the classical JT action \cite{Jackiw:1984je,Teitelboim:1983ux} with a Gibbons-Hawking-York (GHY) term, such that the variational principle is well posed  for spacetimes~$M$ with boundary $B$, and a local counterterm to regulate divergences at $B$
\beq 
\begin{split}
&I_{\text{JT}}^{\text{bulk}}=\frac{1}{16\pi G}\int_{M}\hspace{-2mm}d^{2}x\sqrt{-g}\left((\phi_{0}+\phi)R+\frac{2\phi}{L^{2}}\right),\\
&I^{\text{GHY}}_{\text{JT}}+I^{\text{ct}}_{\text{JT}}=\frac{1}{8\pi G}\int_{B}\hspace{-2mm}dt\sqrt{-\gamma}\left((\phi_{0}+\phi)K-\frac{\phi}{L}\right).
\end{split}
\label{eq:JTclassact}\eeq
Here $\phi$ is the dilaton arising from a spherical reduction of the parent theory, $\phi_{0}$ a constant proportional to the extremal  entropy of the higher-dimensional black hole system, $L$ is the $\text{AdS}_{2}$ length scale, and $K$ is the trace of the extrinsic curvature of $B$ with induced metric $\gamma_{\mu\nu}$. The semiclassical contribution $I_{\text{Pol}}=I_{\text{Pol}}^{\text{bulk}}+I^{\text{GHY}}_{\text{Pol}}+I_{\text{Pol}}^{\text{ct}}$ is composed of the Polyakov action localized via an auxiliary massless scalar field $\chi$ of central charge $c$, a GHY term and counterterm \cite{Almheiri:2014cka} 
\beq
\begin{split}
&I_{\text{Pol}}^{\text{bulk}}=-\frac{c}{24\pi}\int_{M}\hspace{-2mm}d^{2}x\sqrt{-g}\left[(\nabla\chi)^{2}+\chi R\right],\\
&I^{\text{GHY}}_{\text{Pol}}+I^{\text{ct}}_{\text{Pol}}=-\frac{c}{12\pi}\int_{B'}\hspace{-2mm}dt\sqrt{-\gamma}\left(\chi K+\frac{1}{2L}\right).
\end{split}
\label{eq:polyakovact}\eeq
Here $B'$ represents a   cutoff surface near infinity in the flat space regions, which are sewn at the $\text{AdS}_{2}$ boundary~$B$  with transparent boundary conditions, as in \cite{Almheiri:2019yqk}. Notably, the Polyakov action arises as the quantum effective action associated to the conformal anomaly of a $\text{CFT}_{2}$  coupled to any 2D theory of gravity \cite{Christensen:1977jc}.

Semiclassical JT gravity  admits eternal $\text{AdS}_{2}$ black holes as a solution, for which the line element of the metric $g_{\mu\nu}$ in Schwarzschild coordinates $(t,r)$ is 
\beq d\ell^{2}=-f(r)dt^{2}+f^{-1}(r)dr^{2},\quad f(r)=\frac{r^{2}}{L^{2}}-\mu,\label{eq:Schwmet}\eeq
with $\mu$ a mass parameter in the classical Arnowitt-Deser-Misner (ADM) energy. The horizon is at $r_{H}=L\sqrt{\mu}$. 
The remaining semiclassical equations of motion may be solved once the state of the quantum matter is specified.  Requiring regularity at the horizon fixes the vacuum state to be the      Hartle-Hawking  state $|\text{HH}\rangle$ \cite{Hartle:1976tp,Spradlin:1999bn},   for which observers in (null) static  coordinates $(u,v)=(t-r_{\ast},t+r_{\ast})$, with tortoise coordinate $r_{\ast}\in[-\infty,0]$,
see a thermal bath of particles at Hawking temperature $T_{\text{H}}=\frac{\sqrt{\mu}}{2\pi L}$. The expectation value of the normal-ordered  Polyakov stress  tensor   is $\langle \text{HH}|:T^{\chi}_{uu}:|\text{HH}\rangle=\frac{c\pi}{12} T_{\text{H}}^{2}$ (and similarly for $T_{vv}^\chi$). 
With respect to $|\text{HH}\rangle$, the semiclassical correction to $\phi$ is just a constant   \cite{Almheiri:2014cka,Pedraza:2021cvx}
\beq \phi(r)=\phi_{r}\frac{r}{L}+\frac{Gc}{3}= \phi_r \sqrt{\mu} \coth \! \left(\!- \frac{\sqrt{\mu}}{L} r_* \!\right)  + \frac{Gc}{3}\;,\label{eq:dilatonsol}\eeq
where the dimensionless parameter $\phi_{r}>0$ is the boundary value of $\phi$, such that at a cutoff $r=\epsilon_{c}^{-1}$ near the boundary $\phi\to\frac{\phi_{r}}{\epsilon_{c}L}$. 
Recently, moreover, we demonstrated $\chi$ is generically time dependent, such that in the Hartle-Hawking state $|\text{HH}\rangle$ \cite{Pedraza:2021cvx}
\beq 
\begin{split}
\chi&=-\frac{1}{2}\log\left[4\mu\left(1+\frac{\mu UV}{L^{2}}\right)^{-2}\right]\\
&+k-\log\left[\left(K_{U}-\frac{\sqrt{\mu} U}{L}\right)\left(K_{V}+\frac{\sqrt{\mu}V}{L}\right)\right],
\end{split}
\label{eq:chi5}\eeq
where  $(U,V)=(-\frac{L}{\sqrt{\mu}}e^{-\frac{\sqrt{\mu} u}{L}},\frac{L}{\sqrt{\mu}} e^{\frac{\sqrt{\mu} v}{L}})$ are  dimensionful Kruskal coordinates, 
and $k,K_{U},K_{V}$ are constants. 

Directly from the action (\ref{eq:totscact}), using the Noether charge formalism \cite{Wald:1993nt}, we can derive the Wald entropy $S_{\text{W}}$   
\beq S_{\text{W}}\equiv - 2\pi \epsilon_{\mu \nu} \epsilon_{\rho \sigma} \frac{\partial \mathcal L}{\partial R^{\mu \nu \rho \sigma}} =\frac{1}{4G}(\phi_{0}+\phi)-\frac{c}{6}\chi,\label{eq:waldent}\eeq
which is generally evaluated on a bifurcate  Killing horizon with binormal $\epsilon_{\mu\nu}$. Remarkably, imposing   a Dirichlet boundary condition  on $\chi$,   consistent with the transparent boundary condition at~$B$, the integration constants $k,K_{U},K_{V}$ may be fixed such that the semiclassical correction to $S_{\text{W}}$ is exactly the vN entropy of a single  interval with endpoints   $(U_{1},V_{1})$ and $(U_{2},V_{2})$ of a $\text{CFT}_{2}$ in $|\text{HH}\rangle$  with UV regulators $\delta_1,\delta_2$ \cite{Pedraza:2021cvx}
\begin{align} 
S_{\text{vN}}^{\text{HH}}&=-\frac{c}{6}\chi=\frac{c}{12}\log\left[\frac{16\mu^{2}}{\left(1+\frac{\mu U_{1}V_{1}}{L^{2}}\right)^{2}\left(1+\frac{\mu U_{2}V_{2}}{L^{2}}\right)^{2}}\right] \nonumber\\
&+\frac{c}{12}\log\left[\frac{1}{\delta_{1}^{2}\delta_{2}^{2}}(U_{2}-U_{1})^{2}(V_{2}-V_{1})^{2}\right].
\label{eq:vnentHH2}   \end{align}
We emphasize that this result simply follows from solving $\chi$ with Dirichlet boundary. Consequently, while the Wald entropy is evaluated on some bifurcation point, the solution $\chi$ describes the entropy associated with an interval. 
Thus, in semiclassical JT gravity the Wald entropy~(\ref{eq:waldent}) equals the  generalized  entropy $S_{\text{gen}}$   \cite{Pedraza:2021cvx}
\beq S_{\text{W}}= S_{\text{BH}} + S_{\text{vN}}^{\text{HH}}=S_{\text{gen}},\label{eq:SWisSgen}\eeq
which is  the sum of the Bekenstein-Hawking entropy $S_{\text{BH}} = \frac{1}{4G}(\phi_0 + \phi)$ -- corresponding to the area term in \eqref{eq:QESformula} in higher dimensions -- and the CFT entropy $S_{\text{vN}}^{\text{HH}} $ \eqref{eq:vnentHH2}.  
This resolves the discrepancy  found in \cite{Myers:1994sg} between the Wald entropy and generalized entropy in two dimensions.

A key insight of \cite{Pedraza:2021cvx} is that, in the microcanonical ensemble, the generalized entropy is equal to the microcanonical entropy. Further, $S_{\text{gen}}$ is stationary at fixed energy, $\delta S_{\text{gen}}|_{E}=0$, which has the basic elements of the extremization prescription in the QES formula. This observation suggests, at least for backgrounds with $U(1)$ Killing symmetry, that we may compute the generalized entropy and its extremization directly when working with Euclidean path integrals in the microcanonical ensemble.

\section{Microcanonical action and generalized entropy}

We are interested in deriving the extremization prescription in the QES formula (\ref{eq:QESformula}) from first principles. This was accomplished in \cite{Dong:2017xht} using the replica trick. Now we will show for eternal black holes how to derive the extremization of $S_{\text{gen}}$ from a gravitational Euclidean PI in the microcanonical ensemble.


We consider  the saddle-point approximation to the microcanonical gravitational partition function \cite{Brown:1992bq,Brown:1994su}
\begin{equation}
    W (E_0) = \int  \mathcal  D \psi e^{-I_{\text E}^{\text{mc}}( \psi)} \approx e^{-I_{\text E}^{\text{mc}} (  \psi_0)}\,,
\label{eq:microcanZ}\end{equation}
where the path integral is taken over all dynamical fields $\psi=(g_{\mu\nu},\phi,\chi)$ with fixed energy $E_0$ (specified below). In the saddle-point approximation  the Euclidean microcanonical action $I_{\text E}^{\text{mc}}$   is  evaluated on the solutions $ \psi_0 $ of the semiclassical field equations. We will compute the action $I_{\text E}^{\text{mc}}$ on the entanglement wedge \cite{Czech:2012bh,Wall:2012uf,Headrick:2014cta} of an interval $\Sigma$ in $\text{AdS}_{2}$, which is the domain of dependence  of any achronal surface with boundary $\partial \Sigma$. 
The microcanonical boundary condition specifies an  entanglement wedge at fixed energy $E_{0}$ in the  eternal $\text{AdS}_{2}$ black hole background, where the external $\text{CFT}_{2}$ remains in  the Hartle-Hawking vacuum. 

\begin{figure}[t]
 \hspace{-.75cm} \includegraphics[width=4.6cm]{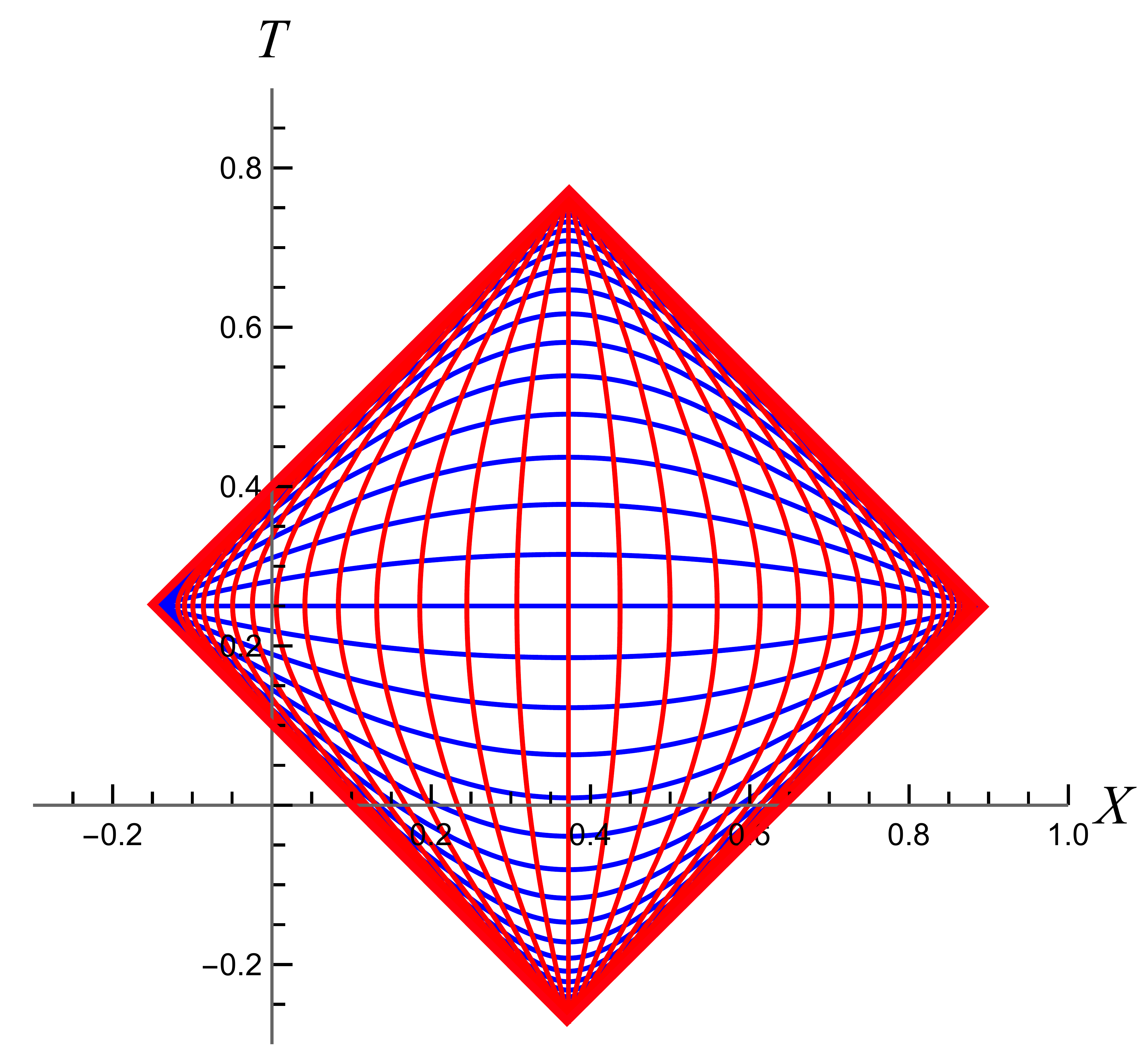}$\quad$\includegraphics[width=4.3
 cm]{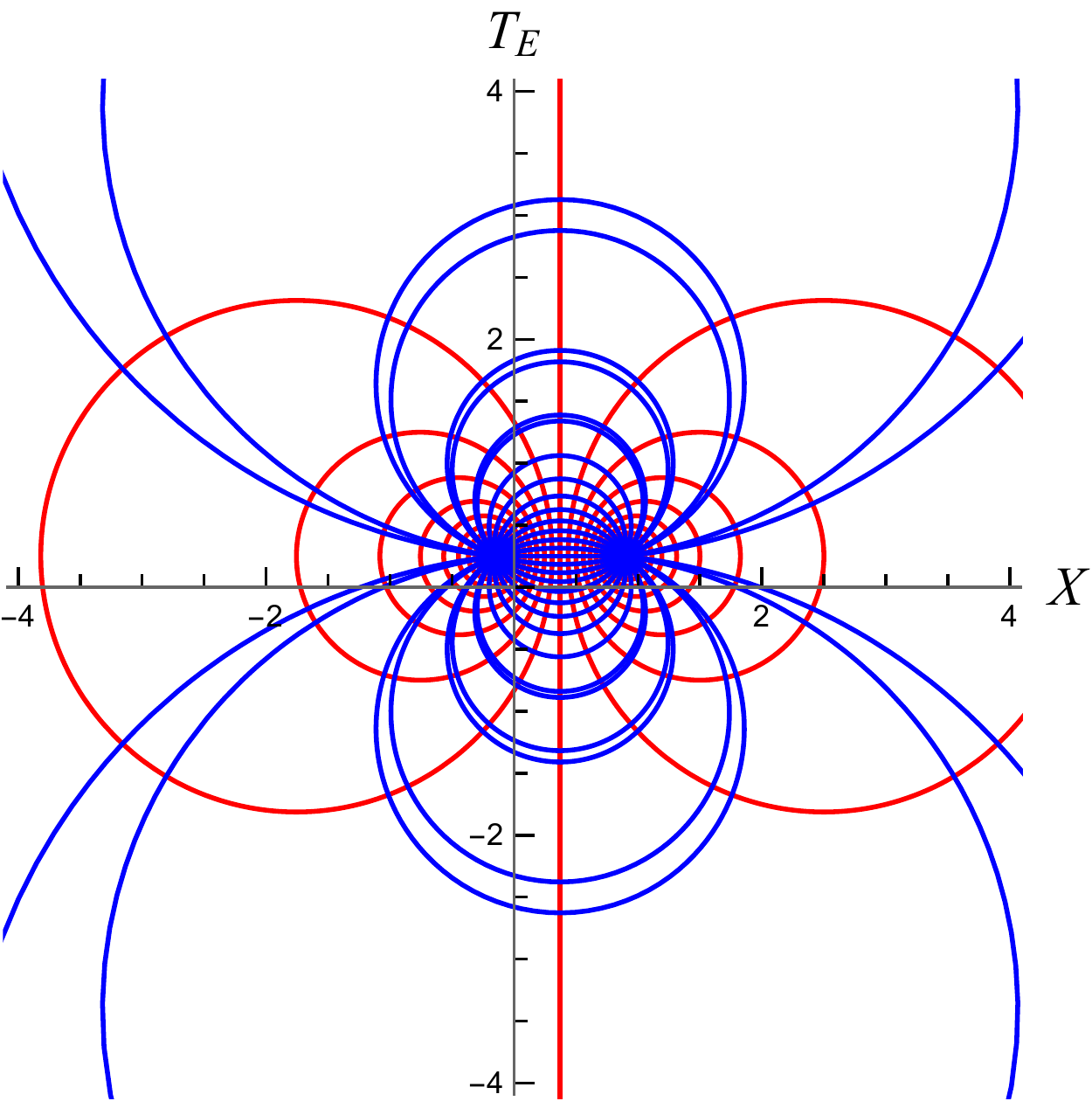}
\caption{Lorentzian $\text{AdS}_{2}$ diamond (left) and its Euclidean continuation (right) in Kruskal coordinates $(T,X)=(\frac{1}{2}(V+U),\frac{1}{2}(V-U))$. Lines of constant $x$ (red) and lines of constant $s,s_{\text{E}}$ (blue) are  at equal  intervals of 0.125. 
On the right,
high contour density corresponds to horizon punctures at $x=\pm\infty$. We have set $a=b=1/2$, $\mu=L=\kappa_{a}=\kappa_{b}=1$ and $u_0\neq v_0\neq 0$.
\label{fig:diamonds}}
\end{figure}
 
 The entanglement wedge consists of a rectangular causal diamond (CD)  with null boundaries    at $(u-u_0=\pm a, v-v_0 =\pm b )$, where $a,b>0$   are in principle different   length scales. Since after extremizing $I_{\text{E}}^{\text{mc}}$ the lengths $a$ and $b$ coincide, we set $a=b$ for purposes of clarity.  Such CDs have a conformal isometry generated by a conformal Killing vector  $\zeta$  \cite{Jacobson:2015hqa,Jacobson:2018ahi,Visser:2019muv}, which we  fix uniquely by demanding $\zeta$ becomes the boost Killing vector of $\text{AdS}_{2}$-Rindler when the future and past vertices of the CD both lie on the AdS boundary (see Appendix \ref{appA}). We cover the CD in  ``diamond universe" coordinates $(s,x)$, adapted to the flow of the vector field~$\zeta$ \cite{Jacobson:2018ahi} (left Fig.~\ref{fig:diamonds}). The coordinate $s$ is the conformal Killing time, satisfying $\zeta \cdot ds =1 $. The   line element in these coordinates is   (Appendix \ref{appB})
  \beq d\ell^{2}=C^{2}(s,x)(-ds^{2}+dx^{2}),\label{eq:diamonduni}\eeq
  with $s,x\in[-\infty,\infty]$ and the conformal factor $C^{2}$ given in~(\ref{eq:conffact}).  
  The  null boundaries  are conformal Killing horizons with constant surface gravity $\kappa$, defined by $\nabla_{\mu}\zeta^{2}=-2\kappa\zeta_{\mu}$ \cite{Jacobson:1993pf}, which is positive on the future horizon and negative on the past horizon (below we take~$\kappa >0$).  When restricting the HH state to the CD, it becomes a thermal density matrix   with temperature $T = \kappa /2\pi$.
  Near the  null boundaries  of the causal diamond at $x=\pm\infty$, the metric  (\ref{eq:diamonduni}) is approximately
  \beq d\ell^{2} \approx  4L^{2}\kappa^{2}e^{\mp2\kappa x}(-ds^{2}+dx^{2}).\label{eq:nearhorizonCD}\eeq
This is simply the    Rindler metric $d \ell^2 = -\kappa^{2}\varrho^{2}ds^{2}+d\varrho^{2}$, with     radial coordinate   $\varrho\equiv 2L e^{\mp\kappa x}$ and  surface gravity $ \kappa=\mp C^{-1}\partial_{x}C |_{x\to\pm\infty}$. Hence, $\zeta = \partial_s$ becomes an approximate boost Killing vector near  $x=\pm\infty$.

Next, we compute the on-shell microcanonical action of CDs in semiclassical JT gravity following the  Hilbert action surface term  method of \cite{Banados:1993qp,Banks:2020tox}. Concretely, we evaluate the GHY term on the boundary of an infinitesimal   disk $D_\epsilon$ of radius $\epsilon$ orthogonal to the  punctures $ \partial \Sigma:x =\pm \infty $ in the   Euclidean   diamond (right Fig. \ref{fig:diamonds}), 
\begin{equation}
    I_{\text{E}}^{\text{mc}} =-\lim_{\epsilon\to 0} \int_{\partial D_\epsilon \times \partial \Sigma}\hspace{-4mm}ds_E \sqrt{ \gamma} K \!\left[\frac{(\phi_0 + \phi)}{8\pi G} - \frac{c\chi}{12 \pi} \right].
\label{eq:microcanactCD}\end{equation}
 Here Euclidean time $s_{\text{E}}=is$ is periodic, $s_{\text{E}}\sim s_{\text{E}}+\frac{2\pi}{\kappa}$, to remove the conical singularity at $x =\pm \infty $ or $\varrho=0$. Note this regularity condition at the horizon is consistent with our choice of the Hartle-Hawking vacuum state \cite{Jacobson:1994fp}. Further, $\sqrt{\gamma} = C$ is the induced metric on   constant $s_{\text{E}}$ slices and  the extrinsic trace  of these slices is $K= \mp C^{-2}\partial_x C$.   Crucially,  in the limit $x\to\pm\infty$, the fields $\phi$ (\ref{eq:dilatonsol}) and $\chi$ (\ref{eq:chi5})  are independent of $s_{\text{E}}$, and $\sqrt{\gamma}K\to\kappa$, leading to
\beq I^{\text{mc}}_{\text{E}}=-S_{\text{gen}} \big |_{\partial \Sigma}, \label{eq:actionentropy}\eeq
where we used \eqref{eq:waldent} and  the nontrivial fact   $S_{\text{W}}=S_{\text{gen}}$.
This is a semiclassical extension of the microcanonical action formula $I_{\text{E}}^{\text{mc}}=-S_{\text{BH}}$ obtained by \cite{Brown:1992bq,Banados:1993qp}, and may be interpreted as a path integral derivation of the generalized entropy of causal diamonds. Note since $\partial \Sigma$  consists of two points, the action is actually twice the entropy.
We further  derive    \eqref{eq:microcanactCD} and  (\ref{eq:actionentropy})   in Appendix \ref{app:Micactfirstlaw} using the Noether charge method \cite{Wald:1993nt,Iyer:1994ys,Iyer:1995kg}, and show that the microcanonical   on-shell action is proportional   to the Noether charge.
Thus,  we find the partition function (\ref{eq:microcanZ}) equals the density of states $W \approx e^{S_{\text{gen}}}$ in the saddle-point approximation, with 
$S_{\text{gen}}$  as the microcanonical entropy.

In   standard thermodynamics the entropy is maximized at fixed energy  in the microcanonical ensemble. Thence, via \eqref{eq:actionentropy}, the microcanonical action $  I^{\text{mc}}_{\text{E}}  $ is minimized at some fixed energy $E_0$. 
For CDs we can infer $E_{0}$  from the variation of the microcanonical action on the full Euclidean diamond $M_{\text E}^{\text{CD}}$, which has a Euclidean time circle $S^1$ of period $2\pi /\kappa$ . Explicitly, 
(see Appendix~\ref{app:Micactfirstlaw}) 
\begin{equation}
    \delta I^{\text{mc}}_{\text{E}}   =  \int_{M_{\text E}^{\text{CD}}} ds_{\text{E}}  \wedge \omega (\psi, \delta \psi, \mathcal L_\zeta \psi) = \int_{S^1} ds_{\text{E}}  \delta H_\zeta \,, \label{eq:varactionmain}
\end{equation}
where   $\omega (\psi,\delta  \psi, \mathcal{L}_{\zeta}\psi)$  is the symplectic current 1-form evaluated on the Lie derivative  $\mathcal{L}_{\zeta}\psi$ of $\psi$ along $\zeta$. In the second equality we inserted Hamilton's equations, $\delta H_\zeta=\int_{\Sigma_{s_{\texttt{E}}}} \omega (\psi, \delta \psi, \mathcal L_\zeta \psi)$, where $  H_\zeta$ is  the  Hamiltonian generating evolution along the flow of $\zeta$ and $\Sigma_{s_{\texttt{E}}}$ are   constant $s_{\text E}$ slices that smoothly intersect  $\partial\Sigma$.  Therefore, the action $I^{\text{mc}}_{\text{E}}$ is stationary at fixed  energy   $E_0\sim H_\zeta$. 
 
 By way of (\ref{eq:actionentropy}), the variational identity (\ref{eq:varactionmain}) is an integrated version of the  semiclassical first law for $\text{AdS}_{2}$ CDs, $\frac{\kappa}{2\pi} \delta S_{\text{W}}=- \delta H_{\zeta}$, which can be derived using the Noether charge method
\cite{Jacobson:2018ahi,Pedraza:2021cvx}. Since the CFT is in the HH state,   $T=\frac{\kappa}{2\pi}$ is identified as the temperature of the causal diamond.
 The first law turns into a proper thermodynamic first law $T \delta S_{\text{gen}} = \delta E_0$ if we identify $E_0 = - H_\zeta.$ Alternatively, one may include the minus sign into the temperature $T$ as done in \cite{Jacobson:2018ahi,Jacobson:2019gco}, however, this  negative temperature seems inconsistent with the thermality of the HH state when reduced to the CD, see   \eqref{eq:diamondthermo}. Moreover,  the first law or \eqref{eq:actionentropy}-(\ref{eq:varactionmain})
tells us the thermodynamic potential defining the microcanonical ensemble is $S_{\text{gen}}(E_{0})$, obeying the equilibrium condition $\delta S_{\text{gen}}|_{E_{0}}=0$. This proves Jacobson's entanglement equilibrium hypothesis  for JT gravity \cite{Jacobson:2015hqa,Bueno:2016gnv,Callebaut:2018xfu}, which thus holds in the microcanonical ensemble. 

\section{Quantum extremal surfaces}

 The extremization prescription in the QES formula (\ref{eq:QESformula}), for eternal AdS$_2$ black holes coupled to a  heat bath, thus arises from extremizing the microcanonical action $I_{\text{E}}^{\text{mc}}$, and will lead to the existence of QESs. Specifically, for $\partial \Sigma$ consisting of one   endpoint in AdS$_2$ and one in flat space near the AdS boundary, the Wald entropy \eqref{eq:waldent} of the latter point vanishes, since there  $\phi=0$ (no gravity) and $\chi=0$ (Dirichlet boundary condition). We can then extremize    $S_{\text{gen}}$ (\ref{eq:SWisSgen}) using (\ref{eq:vnentHH2}) with respect to the first endpoint  $(U_{1},V_{1})$  while holding the second point  $(U_{2},V_{2})$ fixed. Subtracting the extremization conditions   yields $U_1/V_1 = U_2/V_2$ or $t_1 = t_2 $ \cite{Pedraza:2021cvx}. Substituting this into $S_{\text{gen}}$   gives a time-independent result, whereby removing the   divergence near the  boundary ($r_{*,2} \to 0$), yields
\begin{equation}
 S_{\text{gen}} (r_*)=  S_{\text{BH}}(r_*) +\frac{c}{6}\log \left [ 2 \sqrt{\mu} \tanh\left (-\frac{\sqrt{\mu} r_*}{2 L} \right)\right]  
    \end{equation}
with $r_{*}=r_{*,1}.$ 
Extremizing $S_{\text{gen}}$ with respect to $r_*$ leads to  a QES just outside the  classical    horizon \cite{Almheiri:2019yqk,Pedraza:2021cvx}  
\begin{align} \label{eq:QESloc}
r_{\text{QES}} &=\frac{2}{3} r_{H}  \epsilon \sqrt{1 +  \frac{9}{4 \epsilon^2}} \approx r_{H}\left(1+\frac{2\epsilon^{2}}{9}\right), \quad \text{or}\\
 r_{*,\text{QES}} &=-\frac{L}{\sqrt{\mu} } \text{arcsinh}  \left ( \frac{3}{2 \epsilon}\right)  \approx- \frac{L}{\sqrt{\mu}} \left ( \log\left(\frac{3}{\epsilon}\right) +\frac{\epsilon^2}{9}   \right).\nonumber
\end{align}  
The first equality is an exact expression for the QES location, and in the second equality we expanded in terms of the small parameter $\epsilon\equiv\frac{Gc}{\sqrt{\mu}\phi_{r}}\ll1$, which follows  from the semiclassical   regime of validity  $\phi(r_{H})/G\gg c \gg 1$.

\begin{figure}[t]
 \includegraphics[width=\columnwidth]{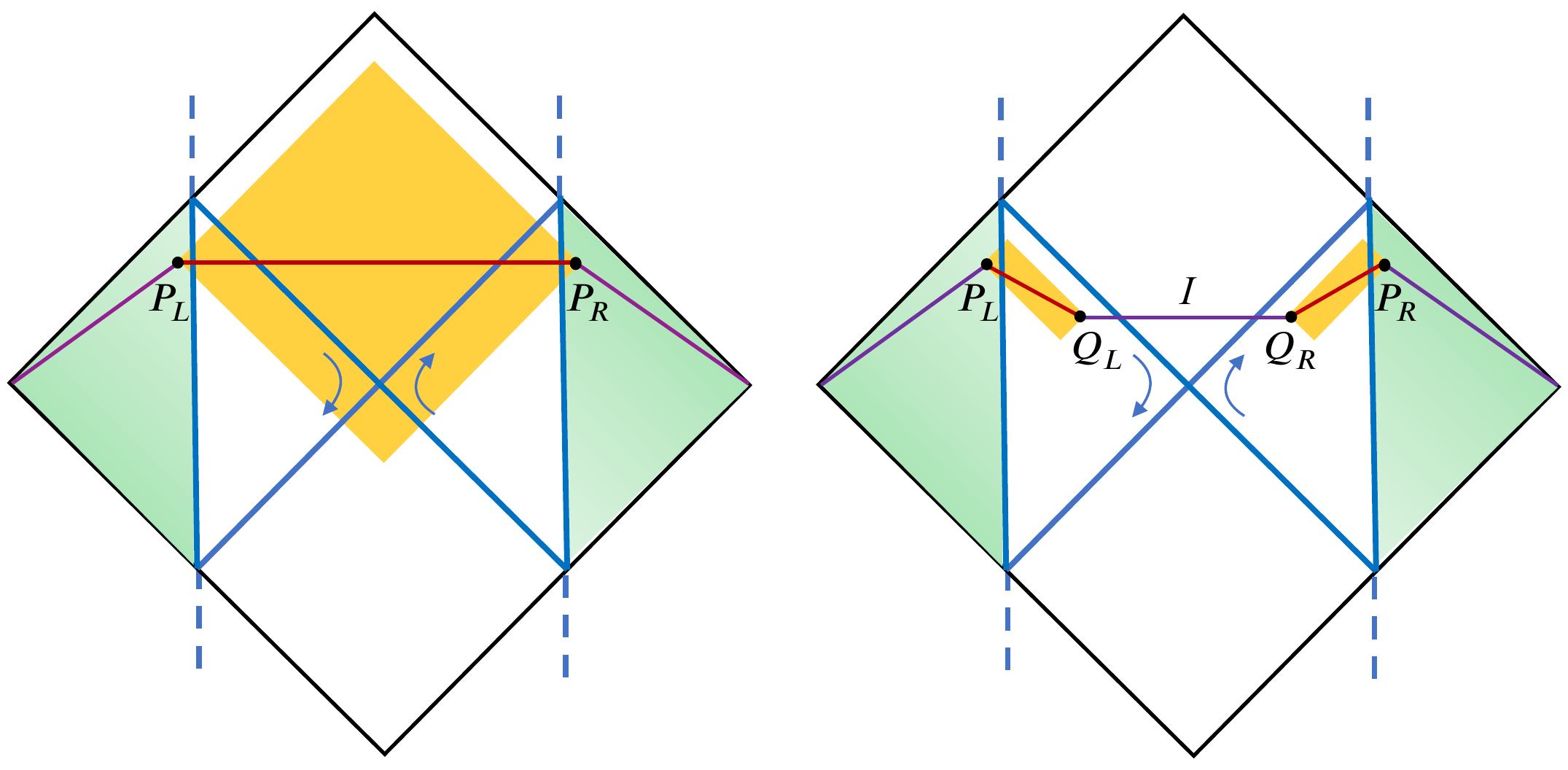}
\caption{Penrose diagram of $\text{AdS}_{2}$ (blue) coupled to a bath (green)  displaying Hawking (left) and island phases (right).  On the right, $Q_{L,R}$ are   quantum extremal surfaces  bounding the island  $I$ (purple). In the Hawking phase the entanglement wedge (yellow) of the black hole  is the Wheeler-deWitt patch and in the island phase it consists of two identical causal diamonds. 
\label{fig:penrosefig}} 
\end{figure}

\section{Islands and entropy of Hawking radiation} 

 We can adapt our prescription to obtain the island and Hawking saddles corresponding to the island and Hawking phases in the Page curve for eternal $\text{AdS}_{2}$ black holes in equilibrium with a flat space bath at temperature $T_{\text{H}}$ \cite{Almheiri:2019yqk} (see Fig. \ref{fig:penrosefig}). Radiation emitted from the black hole into the bath is modeled by a $\text{CFT}_{2}$ at large central charge~$c$, entirely encoded by the semiclassical Polyakov action, and is in the HH state. To maintain equilibrium, radiation entering the baths is compensated by infalling matter, such that the total entropy $S$ of the system is the sum of the entropies $S_{\text{BH}}$ of the two sides of the black hole, $S=2S_{\text{BH}}$. While our derivation above is in Euclidean signature, we analytically continue to Lorentzian signature below.

The island formula technically computes the vN entropy of Hawking radiation using the entanglement wedge of the radiation. However, since the HH state is pure, we instead compute the action $I_{\text E}^{\text{mc}}$ on the entanglement wedge of the black hole, as in \cite{Almheiri:2019qdq,Almheiri:2020cfm}. 
The entanglement wedge of the black hole  is the union of   the domain of dependence  of achronal surfaces with boundaries $\partial \Sigma = \mathcal B \cup P_R$ and  $\partial \Sigma' = \mathcal B'\ \cup P_L$, where  $P_{R,L}$ are   points in the flat region   close to the   boundary and $\mathcal B,\mathcal B'$ are   arbitrary  points in AdS$_2$.
After extremization, $\mathcal B$ becomes the QES $Q_R$ and $\mathcal B' $ turns into $Q_L$ for the island saddle, while $\mathcal{B}'=\mathcal B=P_L$ for the Hawking saddle. Thus, in the island phase the entanglement wedge of the black hole is given by two CDs, while in the Hawking phase the entanglement wedge is a single CD.

In the island phase we compute the on-shell action for the two identical CDs with edges $\partial \Sigma = \mathcal B \cup P_R$  and $\partial \Sigma' = \mathcal B' \cup P_L$ (right Fig. \ref{fig:penrosefig}). We can treat each diamond separately due to large-$c$ factorization \cite{Hartman:2013mia,Rolph:2021nan} or because at times $t\gg t_{\text{P}}\gg \mu^{-1/2}L$    the vN entropy of the two CDs reduces to twice the entropy of one diamond due to an operator product expansion where $P_{R} (P_L)$ and $\mathcal B (\mathcal B')$ are close \cite{Almheiri:2019yqk}.
Focusing on the right CD, extremization of (\ref{eq:actionentropy}) fixes $t_{\mathcal B}=t_{P_{R}}$ (which implies $a=b$) and $\mathcal{B}=Q_{R}$,
such that  $S_{\text{gen}}$  evaluated at the QES (\ref{eq:QESloc}) is \cite{Pedraza:2021cvx}
\beq S_{\text{gen}}(r_{\text{QES}})\approx S_{\text{BH}}  +\frac{c}{6}\log(2\sqrt{\mu})- \frac{c \epsilon}{9}+\mathcal{O}(\epsilon^{2}).
\label{eq:SgenQES}\eeq
Here we used $S_{\text{gen}}|_{\partial \Sigma}=S_{\text{gen}}   |_{Q_R}$ since at $P_{R}$ ($x=\infty$)  we can set $\phi=0$  and $\chi=0$. A similar discussion holds for the left CD.
Including both CDs, since $\phi_{0}/G\gg\phi_{r}\sqrt{\mu}/G\gg c$, for $t>t_{\text{P}}$ we find $S_{\text{vN}}^{\text{rad}}\approx 2S_{\text{BH}}$, consistent with \cite{Almheiri:2019yqk} and up to small corrections near $t\sim t_{\text P}$ \cite{Akers:2020pmf}. Thus, we have derived the constant island phase in the Page curve, where the island is identified as the interval $[Q_L, Q_R]$.


The Hawking phase also follows from extremizing  the action (\ref{eq:actionentropy}). However, since gravity is absent in the flat space bath regions, one neglects the dilaton.
Extremizing $S^{\text{HH}}_{\text{vN}}$ with respect to $(U_{L},V_{L})$ results in   $U_L = V_R$ and $V_L= U_R$ (or, $t_{L}=-t_{R}\equiv- t$ and $r_{\ast,L}=r_{\ast,R}\equiv r_{\ast}$) and $r_{\ast}\ll \mu^{-1/2}L$ and $t\gg \mu^{-1/2}L$. This fixes $\mathcal{B}'=\mathcal{B}=P_{L}$ to be close to the left AdS boundary, such that the Polyakov term in $I^{\text{mc}}_{\text{E}}$ (\ref{eq:microcanactCD}) is evaluated over a Wheeler-deWitt CD (left Fig. \ref{fig:penrosefig}). 
Inserting the extremal points into $S_{\text{vN}}^{\text{HH}}$ yields a time-dependent result~\cite{Almheiri:2019yqk}
\beq 
\begin{split}
S_{\text{vN}}^{\text{HH}}&=\frac{c}{3}\log\left[\frac{2L}{  {\delta}   }\frac{\cosh(\sqrt{\mu}t/L)}{\sinh(-\sqrt{\mu}r_{\ast}/L)}\right]\\
&\approx \frac{c}{3}\frac{\sqrt{\mu}}{L}t+...\,, \quad t\gg L/\sqrt{\mu}.
\end{split}
\label{eq:SvnHHt}\eeq
One may interpret the growth in time    as arising from a sequence of Wheeler-deWitt CDs of each time slice. This linear growth   leads to an information paradox~\cite{Almheiri:2019yqk}. 

Combined, we see extremizing (\ref{eq:actionentropy})   yields the Hawking and island phases of $S_{\text{vN}}^{\text{rad}}(t)$.  
According to the QES formula (\ref{eq:QESformula}) the Page curve follows from a global minimization over the location of the QES, where the turnover between the Hawking and island curves occurs at the Page time $t_{\text{P}}=\frac{6\beta_{\text{H}}}{2\pi c}S_{\text{BH}}\approx\frac{3L\phi_{0}}{2\sqrt{\mu}Gc}\gg1$. This global minimization is not apparent from extremizing $I^{\text{mc}}_\text{E}$ or from the maximization condition on $S_{\text{gen}}$ found here.
This is because we are working in a microcanonical ensemble. Our choice of  ensemble is akin to \cite{Marolf:2018ldl} who found the condition of dominance in the microcanonical path integrals of small eternal AdS black holes corresponds to maximizing the holographic entanglement  entropy \cite{Hubeny:2007xt}. Crucially, maximizing $S_{\text{gen}}$ with respect to the solution at fixed $E_{0}$ is consistent with the QES prescription~\eqref{eq:QESformula} of  extremizing $S_{\text{gen}}$ with respect to the location and shape of $X$ \cite{Marolf:2018ldl}.
Thus, minimizing $I^{\text{mc}}_{\text{E}}$ is  consistent with the QES or island formula.

Note that maximizing $S_{\text{gen}}$ at fixed energy suggests $S_{\text{vN}}^{\text{rad}}(E_0)$ follows a curve mirroring the Page curve, arising from the global minimization of the microcanonical action. That is, for ``low" energies $E_0$, the entropy $S_{\text{gen}}(E_0)$ is maximal for the Hawking saddle, while at ``higher" energies the entropy is maximal for the island saddle. Deriving such a curve, however, entails a more detailed knowledge of the gravitational energy of CDs. 

 
 \section{Discussion}

 The island formula was derived for  eternal and dynamical $\text{AdS}_{2}$ black holes in JT gravity using the ``replica trick" \cite{Dong:2017xht,Almheiri:2019qdq,Penington:2019kki,Goto:2020wnk,Colin-Ellerin:2020mva,Colin-Ellerin:2021jev}. It remains an open question how to generalize the derivation for other theories of gravity, including non-JT 2D dilaton theories of gravity, such as deformed JT \cite{Witten:2020ert}, or 2D flat space analogs \cite{Callan:1992rs,Russo:1992ax,Bose:1995bk}. The microcanonical PI may be able to address this problem. Firstly, the above arguments for the microcanonical PI hold for generic 2D dilaton gravity models with Lagrangian density $\mathcal{L}= L_{0}[Z(\phi)R+U(\phi)(\nabla\phi)^{2}-V(\phi)]+\mathcal{L}_{\text{Pol}}$, for which   the microcanonical on-shell action \eqref{eq:microcanactCD} generalizes to 
 \beq \hspace{-3mm}I^{\text{mc}}_{\text{E}} =-\lim_{\epsilon\to0}\int_{\partial D_{\epsilon}\times \partial\Sigma}\hspace{-6mm} ds_{\text{E}}\sqrt{\gamma}K\left[2L_{0}Z(\phi)-\frac{c\chi}{12\pi}\right].\eeq
Following the derivation above, we again find $I^{\text{mc}}_{\text{E}} = - S_{\text{W}}$ with the Wald entropy $ S_{\text{W}}= 4\pi L_{0}Z(\phi)-c \chi / 6 $. Since the Polyakov contribution may always be cast as the vN entropy, this yields $I^{\text{mc}}_{\text{E}}=-S_{\text{gen}}$ for \emph{any} 2D dilaton gravity theory coupled to conformal matter. Then, applying our arguments here would provide a derivation of  the extremization of $S_{\text{gen}}$ for flat eternal black holes \cite{Gautason:2020tmk,Hartman:2020swn} and 2D de Sitter space \cite{Chen:2020tes,Hartman:2020khs,Balasubramanian:2020xqf,Sybesma:2020fxg}, recently accomplished in \cite{Svesko:2022txo}. This is a distinct advantage over current techniques using the replica trick.
 
 Thus far the island formula has not been derived for theories of gravity in higher dimensions. The so-called Hilbert action boundary term method \cite{Banados:1993qp} or the Noether charge formalism \cite{Iyer:1995kg} employed here holds  for arbitrary theories of gravity in general spacetime dimensions \cite{Iyer:1995kg}. In fact, our derivation of the on-shell microcanonical action in Appendix \ref{app:Micactfirstlaw} is valid for causal diamonds in any theory of gravity. Thus, given the correct equivalent of the semiclassical Polyakov action, the microcanonical path integral may be used, in principle, to derive the island formula for higher-dimensional theories of gravity, and, correspondingly, a Page curve. 
 An obstacle to applying our results to higher dimensions, however, is the fact that our arguments here rely on the equivalence between the Wald entropy and the generalized entropy \cite{Pedraza:2021cvx}. This could be due to the fact that the Polyakov action is 1-loop exact in 2D, capturing the full effect of the conformal anomaly, which might not hold in higher dimensions. 
 
 
 Finally, while the replica trick derives the Page curve of Hawking radiation of dynamical black holes, it is a clear challenge for the microcanonical path integral described here. This is because our results reliably hold in equilibrium since there exists an obvious candidate for (conformal) Killing time, while a dynamical black hole lacks such a symmetry. This problem may be circumvented by instead considering the York time \cite{York:1972sj}, which exists for more general spacetimes, or providing a microcanonical interpretation of dynamical horizon entropy \cite{Iyer:1994ys,Hayward:1998ee,Wall:2015raa}. \\
 


\subsection*{Acknowledgements}

We are grateful to B. Banihashemi, T. Banks, P. Draper, R. Emparan, A. Frassino, F. Gautason, T. Jacobson, D. Marolf, J. Sonner, M. Tomasevic, E. Verlinde and A. Wall for useful discussions. A.S. and M.V. acknowledge the University of Barcelona for hospitality while this work was being completed. J.P. is supported by ``la Caixa'' Foundation (ID 100010434), fellowship code LCF/BQ/PI21/11830029, and by the European Union's Horizon 2020 research and innovation programme under the Marie Sklodowska-Curie grant agreement No 847648. A.S. is supported by the Simons Foundation through \emph{It from Qubit: Simons Collaboration on Quantum Fields, Gravity, and Information}. W.S. is supported by the Icelandic Research Fund (IRF) via a Personal Postdoctoral Fellowship Grant (185371-051). M.V. is supported by  the Swiss National Science Foundation, through Project Grants  200020-182513 and  51NF40-141869 The Mathematics of Physics (SwissMAP).

\appendix

\section{Conformal isometry of \\ causal diamonds in AdS$_2$}
\label{appA}

We first derive the conformal isometry that preserves a causal diamond in a generic two-dimensional spacetime, for a line element in conformal gauge $d\ell^2 =- e^{2\rho}dudv$. Afterwards, we specialize to diamonds in  AdS$_2$. 
Since the property of a conformal isometry is invariant under a Weyl rescaling of the metric, we can leave off the conformal factor  and study the conformal Killing vectors of the line element $dudv$, which take the general form \cite{Jacobson:2015hqa,Jacobson:2018ahi}
\begin{equation}
    \zeta = A(u) \partial_u + B(v) \partial_v\,.
\end{equation}
  Suppose the causal diamond consists of the intersection of the regions $[u-u_0 = -a,u-u_0= a]$ and $[v-v_0 = -b, v-v_0=b].$ The  maximal spatial slice $\Sigma$ is described by $u-u_0 =  - \sqrt{(v-v_0)^2 +a^2 - b^2}$ and the line between the future and past vertices is given by $u-u_0 =   \sqrt{(v-v_0)^2 +a^2 - b^2}$.   The conformal Killing vector that preserves the diamond is only a function of the distance  $u-u_0$ and $v-v_0$, \emph{i.e.}, $A=A(u-u_0)$ and $B=B(v-v_0)$. To map the diamond onto itself $\zeta$ must be tangent to the null generators on the null boundaries $u =u_0 \pm a$ and $v=v_0 \pm b,$ which implies $A(  \pm a)=0$ and $B(  \pm b)=0$. Hence, $A(y)=g_{a}(y)[h(a)-h(y)]$ and $B(y)= m_{b}(y)[n(b)- n(y)]$, with $h(y)=h(-y)$ and $n(y)=n(-y).$ In addition, the flow of $\zeta$ must respect the reflection symmetries  across the line between the future and past vertices and across $\Sigma$, when $a$ and $b$ are interchanged.  In particular, replacing $u-u_0 \leftrightarrow v-v_0$ and $a\leftrightarrow b$ leaves the vector field unchanged $\zeta \to \zeta$, so $m_{b}(y)=g_{b}(y)$ and $n(y)=h(y)$. Further, if $u-u_0 \leftrightarrow -(v - v_0)$ and $a\leftrightarrow b$  we must have $\zeta \to -\zeta$, yielding $g(y)=g(-y)$. Therefore, the conformal isometry of a  causal  diamond is generated by 
\begin{align} \label{eq:appckv1}
    \zeta &=A_a(u-u_0) \partial_u + A_b(v-v_0) \partial_v \, , \quad \text{with}   \\
   A_a(y)   &= g_{a} (y)[h(a)- h(y)],     \,
   g(y)  = g(-y) , \,  h(y)=h(-y), \nonumber
\end{align}
which holds for a generic two-dimensional spacetime. Further, the past and future null boundaries of the diamond are conformal Killing horizons, since $\zeta$ becomes null on these boundaries. Due to the different length scales $a$ and $b$, there are two (positive) surface gravities 
 \begin{equation}
 	\kappa_{a} =   g_a(a) h'(a) \quad \text{and} \quad \kappa_{b} =g_b(b) h'(b)\,,
 \end{equation}
 defined via $\nabla_\mu \zeta^2 = - 2 \kappa \zeta_\mu$ \cite{Jacobson:1993pf} evaluated on the future null boundaries $u-u_0 =   a$ and $v-v_0=b$, respectively. 

Next we place the causal diamond in AdS$_2$ space, for which the line element  in  null coordinates  $u=t-r_*$ and $v=t+r_*$ reads
\begin{equation} \label{eq:adsmetricinuandv}
    d\ell^2= - e^{2 \rho}dudv\,,\quad e^{2 \rho} = \frac{\mu}{\sinh^2 [\frac{\sqrt{\mu}}{2L} (v-u)]}\,.
\end{equation}
As a special case, we consider a  Rindler wedge in AdS space which 
has the shape of a half causal diamond, and becomes a full diamond when AdS is glued to a Minkowski patch at the conformal boundary.
We require that the conformal isometry of  a generic diamond   becomes  the boost isometry   of AdS-Rindler space, if the future and past vertices of the diamond are both  located on the AdS boundary, \emph{i.e.}, if $b= a$ and $r_{*,0}\equiv\frac{1}{2}(v_0-u_0)=0$ or $t_0 \equiv \frac{1}{2}(v_0 + u_0)= v_0 = u_0$. The boost Killing vector of AdS-Rindler space is \cite{Pedraza:2021cvx}
\begin{align} \label{eq:appboost}
    \xi &=A(u-t_0) \partial_u + A(v-t_0)\partial_v \, , \quad \text{with}   \\
   A(y) &= \frac{L \kappa /\sqrt{\mu}}{\sinh(\sqrt{\mu}a/L)}[\cosh(\sqrt{\mu}a/L)- \cosh(\sqrt{\mu}y/L)] \,,\nonumber
\end{align}
for which the surface gravities coincide $\kappa_{a}=\kappa_{b}=\kappa$. Comparing   \eqref{eq:appckv1} and \eqref{eq:appboost}, we see the requirement $\zeta \to \xi$, as $b\to a$ and $v_0,u_0\to t_0$, restricts the   functions in the conformal Killing vector  \eqref{eq:appckv1}  to be
\begin{equation} \label{eq:functionsckv}
    g_{a}(y)= \frac{L \kappa_{a}/\sqrt{\mu}}{\sinh (\sqrt{\mu}a /L)}\,,\,\, h(y)=\cosh (\sqrt{\mu}y /L)\,,
\end{equation}
and similarly for $g_{b}(y).$ Thus, by the special case of the AdS-Rindler (half) diamond, the conformal isometry of a causal diamond in AdS$_2$ is uniquely fixed to be
\begin{align}
    \zeta &= A_a(u-u_0) \partial_u +A_b(v-v_0) \partial_v\,, \quad \text{with}\\
    A_a(y)&= \frac{L \kappa_a / \sqrt{\mu}}{\sinh(\sqrt{\mu}a/L)} \left[ \cosh (\sqrt{\mu}a/ L) -\cosh (\sqrt{\mu}y/ L)  \right]. \nonumber 
\end{align}
For $a=b$ and $\mu =1$ this is equivalent to the conformal Killing vector of a spherically symmetric causal diamond in higher-dimensional  de Sitter spacetime \cite{Jacobson:2018ahi}.

\section{Diamond universe coordinates in AdS$_2$}
\label{appB}

We can cover a causal diamond with the inextendible  coordinates ($s,x$), introduced in \cite{Jacobson:2018ahi} (see also \cite{Banks:2020tox}). This coordinate system is adapted to the flow of the conformal Killing vector $\zeta$ that preserves the diamond. In particular, the coordinate $s \in [-\infty,\infty]$ is the conformal Killing time, defined as the function that satisfies $\zeta \cdot ds =1$ and, for $a =b$, $s=0$ on the maximal slice $\Sigma$ ($t=t_0$).
Further, the  coordinate $x \in [-\infty, \infty]$   is a spatial coordinate and satisfies $\zeta \cdot dx =0$ and $|dx| = |ds|$ and, for $a =b$ we have $x=0$   at $r_* = r_{*,0}$. It follows from these conditions that the two-dimensional line element in   ``diamond universe'' coordinates is
\begin{equation} \label{eq:diamondmetric1}
    d\ell^2 = C^2 (s,x) (-ds^2 + dx^2)= -C^2 (\bar u, \bar v)d \bar ud\bar v\,.
\end{equation}
where $\bar u = s-x$ and $\bar v = s+x$ are null coordinates. The null boundaries of the  diamond are   at $\bar u =\pm\infty$ ($u-u_0=\pm a$) and $\bar v = \pm \infty$ ($v-v_0 = \pm b$), where the plus signs corresponds to the future horizon and the minus signs to the past horizon. On any constant $x$ slice the vertices are located at $s = \pm \infty$, and on a constant $s$ slice the   two edges are  at $x= \pm \infty$. Furthermore, in these coordinates $\zeta = \partial_s=\partial_{\bar u} +\partial_{\bar v}$, which should be equivalent to the conformal Killing vector in \eqref{eq:appckv1} where the functions are given by \eqref{eq:functionsckv} in AdS$_2$. From the equality of these two expressions for $\zeta$ we obtain the   transformation from the null coordinates $(u,v)$ to   $(\bar u, \bar v)$    
\begin{equation}
\begin{split}  \label{eq:transfdiamond1}
    e^{\kappa_a \bar u} &= \frac{\sinh \left [\frac{\sqrt{\mu}}{2L}(a + u  - u_0)] \right]}{\sinh \left [\frac{\sqrt{\mu}}{2L} (a - u +u_0) \right]}\,, \\      e^{\kappa_b \bar v} &= \frac{\sinh \left [\frac{\sqrt{\mu}}{2L}(b + v  - v_0)] \right]}{\sinh \left [\frac{\sqrt{\mu}}{2L} (b - v +v_0) \right]} \,,
\end{split}
\end{equation}
and the inverse transformation 
\begin{equation}
\begin{split}  \label{eq:transfdiamond2}
    e^{\sqrt{\mu} (u-u_0)/ L}& = \frac{\cosh \left [ (\sqrt{\mu}a /L + \kappa_a \bar u)/2 \right]}{ \cosh \left [ (\sqrt{\mu} a/L - \kappa_a \bar u)/2\right]} \,,\\
      e^{\sqrt{\mu}( v-v_0)/ L}& = \frac{\cosh \left [ (\sqrt{\mu}b /L + \kappa_b \bar v)/2 \right]}{ \cosh \left [ (\sqrt{\mu} b/L - \kappa_b \bar v)/2\right]}\,. 
\end{split}
\end{equation}
Kruskal coordinates  $(T,X)=(\frac{1}{2}(V+U),\frac{1}{2}(V-U))$, with $(U,V)=(-\frac{L}{\sqrt{\mu}}e^{-\frac{\sqrt{\mu} u}{L}},\frac{L}{\sqrt{\mu}} e^{\frac{\sqrt{\mu} v}{L}})$, are convenient to visualize the Euclideanized diamond \cite{Banks:2020tox}.

By comparing the metrics \eqref{eq:adsmetricinuandv} and \eqref{eq:diamondmetric1}, and using the transformation in \eqref{eq:transfdiamond2}, we find    
\begin{align}
  & C^2(\bar u, \bar v) 
  =4 \kappa_a \kappa_b   L^2  \left(e^{\frac{  \sqrt{\mu }}{L}2 a}-1\right) \left(e^{\frac{  \sqrt{\mu }}{L}2 b}-1\right)   \times \nonumber\\
   &\times e^{ \frac{ \sqrt{\mu }  }{ L}2r_{*,0} + \kappa_a \bar u+\kappa_b \bar v } \Big [  \left(e^{\frac{
   \sqrt{\mu }}{L}b}+e^{\kappa_b  \bar v}\right) \left(e^{\frac{   \sqrt{\mu }}{L}a+\kappa_a \bar u}+1\right) \nonumber\\
   &-e^{\frac{\sqrt{\mu }  }{L} 2r_{*,0}} \left(e^{\frac{   \sqrt{\mu }}{L}a}+e^{\kappa_a  \bar u}\right) \left(e^{\frac{   \sqrt{\mu }}{L}b+\kappa_b  \bar v}+1\right)   \Big]^{-2}\,,
\label{eq:conffact}\end{align}
where $r_{*,0}\equiv \frac{1}{2}(v_0-u_0).$ Note  for  $r_{*,0}=0$ and $a=b$ we recover AdS-Rindler space since the conformal factor becomes  
$
    C^2  =  \kappa^2 L^2 /\sinh^2 [\frac{\kappa}{2}(\bar v - \bar u)]  \,,
$
where $\kappa = \kappa_a= \kappa_a.$ 


For a generic diamond there exist two different surface gravities,   associated to the two parts of the future conformal Killing horizons, $\bar u = \infty$ and $\bar v = \infty$, given by
\begin{equation}\kappa_a = - C^{-2} \partial_{\bar u} C^2 |_{\bar u \to \infty}\,, \quad \kappa_b = - C^{-2} \partial_{\bar v} C^2 |_{\bar v \to \infty}\,,
\end{equation}
where we applied the definition $\nabla_\mu \zeta^2=-2\kappa \zeta_\mu$ \cite{Jacobson:1993pf}.
As shown in \cite{Jacobson:2018ahi}, surface gravities   satisfying this definition are constant on a bifurcate conformal Killing horizon, so $\kappa_a$ and $\kappa_b$ are constant. As usual, these surface gravities can be interpreted as the  temperatures corresponding to the $a$ and $b$ portions of the conformal Killing horizon. This is because the Hartle-Hawking state, satisfying $\langle \text{HH}|:T^{\chi}_{  u   u }:|\text{HH}\rangle  =  \frac{c\pi}{12}T_{\text H}^{2}$ with $T_{\text H} = \frac{\sqrt{\mu}}{2\pi L}$, is also thermal with respect to the $(\bar u, \bar v)$ coordinates
\begin{equation}
\begin{aligned}
    \langle \text{HH}|:T^{\chi}_{\bar u  \bar u }:|\text{HH}\rangle &= \left(\frac{d u }{d \bar u}\right)^{2}\frac{c\pi}{12}T_{\text H}^{2} -\frac{c}{24\pi}\{ u, \bar u\}   \\
    &=\frac{c\pi}{12}\left(\frac{\kappa_a}{2\pi}\right)^{2}= \frac{c\pi}{12} T_a^2\;,
    \label{eq:diamondthermo}
\end{aligned}
\end{equation}
where we used the anomalous transformation law for the normal-ordered stress tensor, and   inserted \eqref{eq:transfdiamond2}. A similar result holds for $:T_{\bar v \bar v}^\chi:$, \emph{i.e.}, its expectation value in $|\text{HH}\rangle$ is thermal with temperature $T_b=\kappa_b/2\pi$.  This was already known for the special case of AdS-Rindler \cite{Pedraza:2021cvx}, but here we showed the HH state in AdS$_2$  is thermal with respect to any causal diamond. The temperature of the CD thus seems positive and finite, in contrast to the negative temperature interpretation in \cite{Jacobson:2018ahi,Jacobson:2019gco} and the infinite temperature claim in \cite{Banks:2020tox}.

Near the bifurcation points $x = \pm \infty$ of the conformal Killing horizons of the diamond the surface gravity satisfies  the relation $\nabla_\mu \zeta_\nu = \kappa n_{\mu \nu}$ \cite{Jacobson:2018ahi}, where $n_{\mu\nu}=2 u_{[\mu} n_{\nu]}$ is the outward and future pointing binormal, with $u  =C^{-1} \partial_s$ is the future pointing timelike unit normal and $n   =  \pm C^{-1} \partial_x$ is the outward pointing spacelike unit normal at ${x=\pm \infty}$. The surface gravity can be computed to be $\kappa = \mp \frac{1}{2} C^{-2} \partial_x C^2 |_{x=\pm \infty} = (\kappa_a + \kappa_b)/2$. This expression can also be obtained from the periodicity of the Euclidean time. In the Euclideanized diamond spacetime  the conformal Killing horizon maps to punctures at $x\to\pm\infty$. Near $x\to\pm\infty$ 
the  diamond universe line element (\ref{eq:diamondmetric1}) becomes
\beq d\ell^{2}\approx 4L^{2}\kappa_{a}\kappa_{b}e^{\mp(\kappa_{a}+\kappa_{b})x}(-ds^{2}+dx^{2}),\eeq
which is simply flat Rindler space $d\ell^{2}=-\kappa^{2}\varrho^{2}ds^{2}+d\varrho^{2}$ for the  coordinate $\varrho\equiv 4 L\sqrt{\kappa_{a}\kappa_{b}}(\kappa_{a}+\kappa_{b})^{-1}e^{\mp(\kappa_{a}+\kappa_{b})x/2}$, and identifying the surface gravity $\kappa=(\kappa_{a}+\kappa_{b})/2$. Note the null boundaries $x=\pm\infty$ map to the Rindler horizon $\varrho =0$. Upon Wick rotating $s\to -is_{\text{E}}$, removing the conical singularity in the Euclidean spacetime located at the horizon, has us periodically identify $s_{\text{E}}\sim s_{\text{E}}+2 \pi /\kappa$.


%


\section{Microcanonical action in the \\Noether charge formalism}
\label{app:Micactfirstlaw}

In the main text we obtained the  on-shell   microcanonical action (\ref{eq:actionentropy}) using the Hilbert action surface term  method \cite{Banados:1993qp,Banks:2020tox}. However,   the microcanonical action can be derived using various methods. Brown \cite{Brown:1995su} showed an equivalence between the GHY surface action \cite{Banados:1993qp} and the microcanonical action developed by Brown and York \cite{Brown:1992bq}. Later, both actions were expressed in the Noether charge formalism \cite{Wald:1993nt,Iyer:1995kg}, and shown to be equivalent to the Noether charge. These methods are typically applied to black hole spacetimes, but here we use them to define the microcanonical action for causal diamonds.  Specifically, employing the Noether charge formalism,  below we define the off-shell Euclidean microcanonical action of  CDs, and show it equals   (\ref{eq:microcanactCD}) and  \eqref{eq:actionentropy}  on shell. 

We consider generic semiclassical 2D dilaton gravity theories with   Lagrangian 2-form   $L=\epsilon L_{0}[RZ(\phi)+U(\phi)(\nabla\phi)^{2}-V(\phi)]+L_{\text{Pol}}$, with   $\epsilon$ the spacetime  volume form on the Lorentzian CD spacetime $M^{\text{CD}}$. We foliate  the CD with spacelike slices $\Sigma_s$, labeled by the conformal Killing time $s$, which smoothly intersect the bifurcation points $\partial \Sigma$ of the null boundaries. The Euclidean diamond spacetime $M_{\text E}^{\text{CD}}$ is defined by periodically identifying the Euclidean time $s_{\text{E}}=is$ with $2\pi/\kappa$ to avoid a conical singularity at $\partial \Sigma$. Motivated  by \cite{Brown:1992bq,Iyer:1995kg}, we define the off-shell microcanonical Euclidean  action
\beq
\begin{split}
I^{\text{mc}}_{\text{E}}&\equiv-i \left[\int_{  M_{\text E}^{\text{CD}}}\hspace{-1mm}  L-\int_{M_{\text E}^{\text{CD}}}\hspace{-1mm}ds\wedge \theta(\psi,\mathcal{L}_{\zeta}\psi)\right],
\end{split}
\label{eq:Imicrogen}\eeq
where $\theta$ is the symplectic potential 1-form, and the dynamical fields are $\psi=(g_{\mu\nu},\phi,\chi)$. Note  $\theta(\psi, \mathcal L_\zeta \psi) $ here is non-vanishing since $\zeta$ is a conformal Killing vector instead of a Killing vector. Writing $L = ds \wedge \zeta \cdot L$, we see the two terms between brackets combine into an integral over  the   Noether current 1-form $j_{\zeta}\equiv\theta(\psi,\mathcal{L}_{\zeta}\psi)-\zeta\cdot L$ associated with diffeomorphisms generated by~$\zeta$. 
Using the on-shell identity $j_{\zeta}=dQ_{\zeta}$, with $Q_{\zeta}$ the Noether charge 0-form, and applying Stokes' theorem we find  the  on-shell Euclidean microcanonical action for CDs is equal to 
\begin{equation}
     I^{\text{mc}}_{\text{E}} = \int_{\partial M_{\text E}^{\text{CD}}} ds_{\text{E}} \wedge Q_\zeta,
\end{equation} 
where the total Noether charge for semiclassical 2D dilaton gravity 
is $Q_\zeta = Q_\zeta^{\phi}  + Q_\zeta^{\text{Pol}}$
with \cite{Pedraza:2021cvx}
\begin{align}
    Q_\zeta^{\phi}   &=-L_{0}\epsilon_{\mu\nu}[ Z(\phi)\nabla^\mu \zeta^\nu + 2 \zeta^\mu \nabla^\nu Z(\phi)] , \nonumber \\
     Q_\zeta^{\text{Pol}}   &=\frac{c}{24\pi} \epsilon_{\mu\nu}[ \chi \nabla^\mu \zeta^\nu + 2 \zeta^\mu \nabla^\nu \chi].
\end{align} 
where $\epsilon_{\mu \nu}=-n_{\mu\nu}$ is the binormal volume form on   $\partial \Sigma$,   with $n_{\mu\nu}$ as the outward and future pointing binormal. Since $\partial M_{\text E}^{\text{CD}}$ has topology $S^1 \times \partial \Sigma$ we restrict the Noether charge to $\partial \Sigma    $, where $\zeta |_{\partial \Sigma}=0$ and $\nabla_\mu \zeta_\nu |_{\partial \Sigma}= \kappa n_{\mu \nu}$, hence it becomes
$
    Q_\zeta |_{\partial \Sigma} = - \frac{\kappa}{2\pi}\left [  4\pi L_{0}Z(\phi) - \frac{c}{6} \chi\right]
$. 
Importantly, 
  $Q_\zeta |_{\partial \Sigma} $ is independent from $s_{\text{E}}$, because $\phi, \chi$ are constant   in the limit $x \to \pm\infty.$ Thus, we can  integrate out the Euclidean time, and obtain that the on-shell microcanonical action is equal to minus the Wald entropy   
\begin{equation}
    I^{\text{mc}}_{\text{E}} =   \frac{2\pi}{\kappa}\oint_{\partial \Sigma} Q_\zeta =   - S_{\text{W}} \big |_{ \partial \Sigma}\,, \label{eq:microaction2}
\end{equation}
where $ S_{\text{W}}= 4 \pi L_{0}Z(\phi)-c \chi / 6 $. This shows the microcanonical action is proportional to the Noether charge and it proves \eqref{eq:actionentropy}, given $S_{\text{W}}=S_{\text{gen}}$. 

In a proper microcanonical ensemble the action is minimized at fixed energy $E_0$. The definition of $E_0$ in our setup  may  therefore be identified from the variation of the on-shell action on $M_{\text E}^{\text{CD}}$
\begin{equation}
    \delta I^{\text{mc}}_{\text{E}}   =  \int_{M_{\text E}^{\text{CD}}} ds_{\text{E}}  \wedge \omega (\psi, \delta \psi, \mathcal L_\zeta \psi) = \int_{S^1} ds_{\text{E}}  \delta H_\zeta \,, \label{eq:varaction}
\end{equation}
 where the symplectic current 1-form is    defined as $\omega (\psi,\delta_1 \psi, \delta_2 \psi) \equiv  \delta_1 \theta (\psi,\delta_2 \psi) - \delta_2 \theta (\psi,\delta_1 \psi) $, and in the second equality we inserted $\delta H_\zeta=\int_{\Sigma_{s_{\texttt{E}}}}\omega (\psi, \delta \psi, \mathcal L_\zeta \psi)$, the variation of the Hamiltonian generating evolution along the flow of $\zeta$. The first equality follows from varying \eqref{eq:Imicrogen} and using $\delta L= E^\psi \delta \psi + d \theta (\psi,\delta \psi)$, where $E^\psi=0$ are the field equations for $\psi$, and inserting  Cartan's magic formula $\zeta \cdot d \theta = \mathcal L_\zeta \theta - d (\zeta \cdot \theta)$. The term $\int_{\partial M_{\text E}^{\text{CD}}} ds \wedge \zeta \cdot \theta$ vanishes because $\zeta$ is zero at the edge~$\partial \Sigma$. Thus, we see from  \eqref{eq:varaction}    the action $I^{\text{mc}}_{\text{E}}$ is minimized at fixed  energy   $E_0\sim H_\zeta$ up to a sign and constant. We set the constant to zero and  the sign is fixed by comparing to the first law $\frac{\kappa}{2\pi} \delta S_{\text{W}}=-\delta H_{\zeta}$ such that $E_0=-H_{\zeta}$.

 Lastly, let us comment on the equivalence between the on-shell microcanonical action (\ref{eq:microaction2}) and the GHY surface term used in the main text
 \beq I^{\text{mc}}_{\text{E}}=-\lim_{\epsilon\to0}\int_{\partial D_{\epsilon}\times \mathcal{H}}\hspace{-4mm} ds_{\text{E}}\sqrt{\gamma}K\left[2L_{0}Z(\phi)-\frac{c\chi}{12\pi}\right].\label{eq:microcanactCD2}\eeq
where $D_\epsilon$ is an infinitesimal disk of radius $\epsilon$ orthogonal to
the    punctures $\partial \Sigma : x =\pm \infty$
in the Euclideanized diamond. While initially obscure, the equivalence of (\ref{eq:microaction2}) and (\ref{eq:microcanactCD2}) follows from the fact that, on-shell, $H_{\zeta}=0$ on the bifurcation surface and that the GHY boundary term is independent of $s$ on $\partial \Sigma$ \cite{Iyer:1995kg}. More explicitly, the Hamiltonian $H_{\zeta}$ for a theory which fixes the induced metric of the boundary $\partial M$ of a (Lorentzian) manifold $M$ is generically given by an integral over the codimension-2 slices $\mathcal{C}_{s}$ where $\Sigma_{s}$ orthogonally intersects $\partial M$ \cite{Pedraza:2021cvx} (see also \cite{Iyer:1995kg,Harlow:2019yfa})
\beq H_{\zeta}=\int_{\mathcal{C}_{s}}(Q_{\zeta}-\zeta\cdot b)= \oint_{\mathcal{C}_{s}}\epsilon_{\partial\Sigma} N\varepsilon.\eeq
Here $b=\epsilon_{B}K[-2L_{0}Z(\phi)+\frac{c}{12\pi}\chi]$ is the GHY boundary term 1-form,  $\varepsilon=-2L_0 n^\alpha \nabla_\alpha Z(\phi)$ is the quasi-local energy density, and $N=-\zeta^\mu  u_\mu$ is the lapse. Crucially, on bifurcation points $\partial\Sigma$, the lapse $N=0$ such that $H_{\zeta}=0$. 
Then, for a 1-parameter family of surfaces $(\partial \Sigma)_{\epsilon}$ in $\Sigma_{s_{\text{E}}}$, where $\lim_{\epsilon\to0}(\partial \Sigma)_{\epsilon}\to \partial \Sigma$, we have
\beq \lim_{\epsilon\to0}\int_{(\partial \Sigma)_{\epsilon}}Q_{\zeta}=\lim_{\epsilon\to0}\int_{(\partial \Sigma)_{\epsilon}}\zeta\cdot b.\eeq
Thus, by the  definition $S_{\text{W}}=-\frac{2\pi}{\kappa}\int_{\partial \Sigma }Q_{\zeta}$ of the Wald entropy,   we find 
\begin{align}
 S_{\text{W}}&=-\lim_{\epsilon\to0}\frac{2\pi}{\kappa}\int_{(\partial \Sigma)_{\epsilon}}\zeta\cdot b=-\lim_{\epsilon\to0}\int_{\partial D_{\epsilon}\times \partial \Sigma}\hspace{-4mm}ds_{E}\wedge\zeta\cdot b \nonumber \\
 &=\lim_{\epsilon\to0}\int_{\partial D_{\epsilon}\times \partial \Sigma}\hspace{-4mm} ds_{E}\sqrt{\gamma}K\left[2L_{0}Z(\phi)-\frac{c\chi}{12\pi}\right],
\end{align}
 where $2\pi/\kappa$ was replaced with the integral $\int_{\partial D_{\epsilon}}ds_{\text{E}}$ and we used that the GHY term $b$ is independent of $s_{\text{E}}$ on $(\partial \Sigma)_{\epsilon}$. This shows (\ref{eq:microaction2}) and (\ref{eq:microcanactCD2}) are equal, which establishes the Hilbert  action surface formula \eqref{eq:microcanactCD}.

\bibliography{nowormshortrefs}

\end{document}